\documentclass[12pt,preprint]{aastex}

\def\msun{M$_\odot$}

\def\cm3{cm$^{-3}$}
\def\kms{km\thinspace s$^{-1}$}

\def\ni{\noindent}

\def\13co{$^{13}$CO}
\def\ceio{C$^{18}$O\ }

\def\ctfs{C$^{34}$S\ }

\def\ntwohp{N$_2$H$^+$\ }
\def\ntwohpp{N$_2$H$^+$}
\slugcomment{To appear in the Astrophysical Journal}

\shorttitle{Dynamical State of B68}
\shortauthors{Lada et al.}

\begin{document}


\title{The Dynamical State of Barnard 68:\\
A Thermally Supported, Pulsating Dark Cloud.}


\author{Charles J. Lada, and Edwin.A. Bergin}
\affil{Harvard-Smithsonian Center for Astrophysics, 60 Garden Street,
Cambridge MA 02138}
\email{clada@cfa.harvard.edu ebergin@cfa.harvard.edu}
\author{Jo\~ao. F. Alves}
\affil{European Southern Observatory, Garching Germany}
\email{jalves@eso.org}

\and

\author{Tracy L. Huard}
\affil{Harvard-Smithsonian Center for Astrophysics, 60 Garden Street, 
Cambridge, MA. 02138}
\email{thuard@cfa.harvard.edu}


\begin{abstract}
We report sensitive, high resolution molecular-line observations
of the dark cloud Barnard 68 obtained with the IRAM 30-m telescope. 
We analyze spectral-line observations of \ceio(1--0), C$^{32}$S(2--1), \ctfs(2--1),
and \ntwohp(1--0) in order to investigate the kinematics and dynamical state of 
the cloud. We find extremely narrow linewidths in the central  regions
of the cloud, $\Delta V = 0.18 \pm 0.01$ \kms and $0.15 \pm 0.01$ \kms
for \ceio and
\ctfs, respectively. These narrow lines are consistent with thermally broadened
profiles for the measured gas temperature of 10.5 K. We determine the thermal
pressure to be a factor 4 -- 5 times greater than the non-thermal (turbulent) 
pressure in the central regions of the cloud, indicating that
thermal pressure is the  primary source of support against 
gravity in this cloud. This confirms
the inference of a thermally supported cloud drawn previously from
% the striking agreement
%of the cloud's radial density distribution derived from 
deep infrared extinction measurements 
%and the predictions for a critically stable Bonnor-Ebert sphere
\citep{all01}.  We also find the molecular linewidths to systematically 
increase in the outer regions of the cloud, where we calculate the thermal
pressure to be between 1 -- 2 times greater than the turbulent pressure. 
We find the distribution of line-center radial velocities for both \ceio and \ntwohp 
to be characterized by systematic and well-defined linear gradients across 
the face of the cloud. The rotational kinetic energy is found to be only
a few percent of the gravitational potential energy, indicating that the contribution
of rotation to the overall stability of the cloud is insignificant. However,
the \ceio and \ntwohp velocity gradients differ from each other in both
magnitude and direction, suggesting that the cloud is differentially rotating,
with the inner regions rotating slightly more slowly than the outer regions. Finally,
our observations show that C$^{32}$S line is optically thick and self-reversed
across nearly the entire projected surface of the cloud. The shapes of 
the self-reversed profiles are asymmetric and are found to vary across
the cloud in such a manner that the presence of 
both inward and outward motions are observed within the cloud. Moreover, 
these motions appear to be globally organized in a clear and systematic alternating spatial 
pattern which is
suggestive of a small amplitude, non-radial oscillation or pulsation of 
the outer layers of the cloud about an equilibrium configuration.

\end{abstract}



\section{Introduction}

Barnard 68 is a fine example of a small round and optically opaque dark
molecular cloud known as Bok Globule.  Recently \citet[hereafter ALL01,]{all01}
%Alves, Lada \& Lada (2001; here
%forward ALL01) 
used deep near-infrared extinction measurements to map the
structure of this globule in a detail previously unsurpassed for any molecular
cloud.  In particular, they were able to construct a highly resolved radial
density profile of the cloud, spanning its entire 100 arc sec (10$^4$AU) extent
with 5 arc sec (500 AU) angular resolution.  They found the measured radial
density distribution of the globule to be in impressive agreement with the
theoretical predictions for a Bonnor-Ebert sphere, that is, a truncated
isothermal sphere bounded by a fixed external pressure.  

The extremely close agreement between observation and theory for this cloud
implies a number of interesting observable consequences (or predictions) for its
physical state.  In particular, thermal pressure is likely to be an important,
if not dominant, source of support for the cloud against gravitational collapse.
In this sense, the physical state of B68 would not be typical, since turbulence
is the dominant source of internal bulk motions and support in most molecular
clouds.  Further evidence in support of the idea that thermal, not turbulent,
motions dominate the internal pressure of the cloud followed from a more detailed
analysis of the extinction data.  Alves, Lada \& Lada (2002) found the surface
density structure of the cloud to be extremely smooth placing a limit on the
magnitude of random fluctuations in the surface density of $\leq$ 5\% over
angular scales ranging from $\sim$ 3 -- 200 arc sec ( 300 -- 20,000 AU).  Such
smooth structure seems incompatible with expectations for turbulent cloud models
 \citep{pbn97, juvela98}.  An important observational consequence for a
thermally dominated cloud would be the presence of very narrow molecular-line
widths.  One might also expect that random variations in molecular-line center
velocities (of optically thin species) across the cloud would be small compared
to the sound speed in the cloud. Both of these aspects of a cloud's velocity
field can be directly investigated with sensitive, high frequency resolution
molecular-line observations.

To further examine and better constrain the physical nature of this unusual
cloud we have obtained extensive, high angular resolution and high
signal-to-noise observations of B68 in a number of interesting molecular
species, including C$^{18}$O, C$^{32}$S, C$^{34}$S, and N$_2$H$^+$.  These
observations were designed, in part, to assess the relative roles of turbulence
and thermal pressures for the support of this cloud against gravity.  Because
B68 was found to be formally unstable and near the critical condition or
pivotal point for gravitational collapse (ALL01), we also designed our
observational program to measure the cloud's dynamical state (i.e., is B68
rotating, expanding, collapsing or static?).

\section{Telescope, Spectrometer and Receivers}

We used the 30m IRAM millimeter-wave telescope located at Pico Valeta
in Spain for the observations reported here. Observations were obtained
during three periods: 2000 April and August and 2001 April.
The dual mixer, dual-channel
receiver was tuned to observe the J = 1--0 transition of \ceio
at 109.78218 GHz, the 1--0 transition of \ntwohp at 93.173178 GHz,
the J = 2--1 transitions of C$^{32}$S at 97.980968 GHz and \ctfs at 
96.412982 GHz. Observations of the J= 1--0 transition of C$^{17}$O and
the J = 3--2 transition of \ntwohp were also obtained, and reported
elsewhere \cite{bahl02}. The half-power beamwidth provided by the telescope 
was 22", 25", and 26" for CO, CS and \ntwohpp, respectively.
System temperatures were typically in the range of 160 -- 190 K.
Observations of the various molecular transitions were acquired in the 
frequency-switched mode.  
The spectrometer was an autocorrelator
configured to provide velocity resolutions of 0.053 \kms for \ceio,
0.06 \kms for \ctfs, 0.137 \kms for \ntwohpp, and 0.03 \kms 
for C$^{32}$S observations.

\section{Observational Results}

\subsection{Integrated Intensity Maps}

We have made integrated intensity maps of C$^{32}$S, 
\ceio and \ntwohp across the B68 cloud. These maps
are displayed in Figure 1. The field that was mapped in CO and CS 
encompassed a larger area than that mapped in \ntwohpp.  The CO observations 
were sampled at the Nyquist rate. The CS and 
\ntwohp observations were obtained with Nyquist sampling in the
inner 50" of B68 and with full beam honeycomb sampling in the outer part
of the map. The CO and
CS distributions show very different morphology, with the CS emission 
peaking in the southeast extension or tail of B68. The \ntwohp emission
has a distribution which most closely traces the column density of dust
and presumably therefore H$_2$ \citep{all01,bahl02}. These differences in spatial distributions
are most likely due to differential depletion of the respective molecular
species as described in detail in our earlier paper \citep{bahl02}.

\subsection{Velocity Field}

Figure 2 shows the individual spectra of  C$^{32}$S, \ctfs, \ntwohp and \ceio lines
observed at the center of our mapping grid. The \ceio and \ctfs
profiles are all single-peaked and characterized by symmetric, Gaussian-like 
shapes which peak at the same velocity. These lines 
are characterized by small optical depths \citep[see,][]{bahl02}. 
The N$_2$H$^+$ which has modest
opacity ( $\tau \approx 2$) also appears single peaked, but exhibits a slight
asymmetry in this high resolution spectrum. This asymmetry in the line 
profile is unresolved in the lower velocity resolution (0.137 \kms) 
observations used to map the emission across the globule. The \ntwohp line also
appears to be centered at a slightly different velocity than the optically
thin CO and CS lines, but this likely reflects the uncertainty of the 
rest frequency of \ntwohpp. On the other hand, the C$^{32}$S 
(hereafter CS) line is double-peaked, suggesting high line center opacity 
and self-reversed line structure. To probe the velocity field and kinematics of B68
we have mapped the line center velocity and linewidth distributions of both
\ceio and \ntwohpp, the two single-peaked lines,
as well as the velocity distribution of the CS self-reversal across the cloud.
The results are described in the following sections.

\subsubsection{Linewidth Distribution}

The linewidths of the observed molecular transitions in B68
are generally quite small and the narrowest lines have
widths very close to that expected from purely thermal Doppler broadening.
Figure 3 shows the spatial distributions of \ceio and \ntwohp
linewidth across
the cloud. The linewidths were determined from gaussian fits to
the observed profiles. In the case of \ntwohp the linewidths were
determined from a simultaneous fitting of all the hyperfine components.
In both maps there appears to be a systematic increase of 
linewidth with distance from the map center in the
northwest and southeast directions. In both tracers the minimum linewidths 
are found in an elongated (NE-SW) region passing through the center of
each map. We plot in Figure 4 the radially averaged variation of 
\ceio and \ntwohp linewidth as a function of radial distance from the 
center (0,0 position) of the cloud maps. The plots show a slow rise in linewidth
with distance in the inner 60 arc sec of the globule. This rise is
followed by a relatively flat variation of linewidth with distance 
until a projected radius of about 100 arc sec, which marks the extent of the 
main B68 cloud. For 
larger radii, where only \ceio observations exist, the
linewidths appear to rapidly increase with radius. These larger distances
mostly correspond to the extended tail region of B68, where the brightest
CS emission is observed. It is interesting that the \ntwohp lines are
everywhere wider than the \ceio lines since chemical considerations
indicate that the two species originate in different 
cloud volumes with \ntwohp interior to \ceio \citep{bahl02}. Radiative
transfer models for the cloud indicate that the higher
opacity of the \ntwohp lines can account for some, but probably not all, of this
difference (Bergin et al 2002).  

\subsubsection{Radial Velocity Distribution}

Figure 5 shows the distribution of line center radial velocity
for \ceio and \ntwohp across B68. The line center velocities were
derived from Gaussian fits to the observed line profiles. For \ntwohp
the line center velocities were derived from 
simultaneous fits to all the hyperfine components.
Both maps show what appears to be
a systematic, well-behaved velocity gradient across the cloud's face.
In the \ceio velocity map, the lines of constant velocity are very
nearly parallel to each other and are aligned roughly in a north-south
direction. The \ntwohp map covers a smaller region and also exhibits
an east-west velocity gradient, although of apparently smaller magnitude.
Both velocity gradients are consistent with that expected for solid body
rotation. 
Assuming that the velocity gradient is linear when projected
on the sky, as would be the case for solid body rotation,
the observed velocity at any point on the projected surface of 
the cloud is related to the velocity gradient and given by:

\begin{equation}
v_{lsr} = v_0 + {dv \over ds}  \Delta \alpha cos \theta
 + {dv \over ds} \Delta \delta sin \theta 
\end{equation}
 
\ni
 Here
$\Delta \alpha$ and $\Delta \delta$ are the offsets in right ascension
and declination expressed in seconds of arc, ${dv \over ds}$ is the 
magnitude of the velocity gradient in the plane of the sky, $\theta$ is its 
direction (as measured from east of north), and $v_0$ 
is the systemic radial velocity of the cloud. To determine the magnitude
of the velocity gradient, its direction on the sky and the 
cloud's systemic velocity, 
we have performed a least squares fit of a two dimensional plane to 
the observed radial velocity 
distributions of \ceio and \ntwohpp, similar to the procedure described
by \citet{gbfm93}. 
For the \ceio map we find the fitted parameters to be: 
${dv}\over{ds}$ = 2.25 $\pm$ 0.024 m s$^{-1}$ arcsec$^{-1}$
 (4.77 \kms pc$^{-1}$ at 100 pc),
V$_{LSR}$ = 3.3614 $\pm$ 0.0009 \kms and $\theta$ = 5.5 $\pm$ 1.2$^o$.
For \ntwohp these parameters are found to be:
${dv}\over{ds}$ = 1.72 $\pm$ 0.09 m s$^{-1}$ arcsec$^{-1}$
(3.52 \kms pc$^{-1}$
at 100 pc), V$_{LSR}$ = 3.3722 $\pm$ 0.0018 \kms and $\theta$ = 23.5 $\pm$
1.2$^o$. The quoted uncertainties are 3 $\sigma$ statistical
uncertainties. Both the magnitude and
direction of the velocity gradients in the \ceio and \ntwohp maps differ
from each other. Differences in these quantities could perhaps arise due
to the opacity difference in the two species. Higher velocity resolution
mapping observations of \ntwohp would be needed to confidently evaluate this
possiblilty. However, the close agreement
of the derived cloud systemic velocity from both species suggests
that such opacity related differences may be small.

\subsubsection{Self-Reversed CS Velocity Field}

The double-peaked CS line is characteristic of  self-reversed
optically thick emission. The fact that the \ctfs line is single
peaked and symmetric confirms the optically thick, self-reversed
nature of the main CS line. At the center of the map, the CS profile
displays the classic red-shifted asymmetry usually associated with
infall motions of the outer cloud material \citep[e.g.,][]{wlymw86,mmtww96}.
This is likely to be the correct interpretation for the gas motions
at the projected center of B68, since the excitation temperature of the lines likely 
increases inwards as a result of the steep, inward increasing 
density gradient (ALL01).
Self-reversed CS lines are found  across nearly the entire extent 
of the B68 cloud. Figure 6 displays a series of CS spectra along
an east-west cut 24 arc sec south of the center of the mapping grid. Although
the self-reversal in the CS profiles is clearly spatially extended, 
the sense of the reversal, or line asymmetry, flips and changes sign
both west and east of the map center. In order to better ascertain 
the nature of the spatial variation in the CS line shapes, we have
compared the velocity of the strongest peak in a CS self-reversed
line profile  with the line center velocity of the optically
thin \ceio line emission at each position in the CS map of B68.
Figure 7 displays a map of the velocity difference, $\delta V $ =
v(\ceio) -- v(CS), that  we constructed from gaussian fitting of the 
CS and \ceio line profiles of this cloud. The differences range between
roughly -0.16 and +0.12 km s$^{-1}$. These values are less than
the corresponding linewidths at each position and also everywhere subsonic.
A positive value
of $\delta V$ indicates a profile with a stronger blue peak and
a red-shifted or infall asymmetry, while
a negative value corresponds to a profile with a stronger red peak 
and a blue-shifted or outflow asymmetry. The most significant feature
of the map is the striking spatial segregation of the blue and red
shifted asymmetry. The red-shifted inflow asymmetry dominates the
central regions of the cloud map but is almost entirely surrounded by
a contiguous region dominated by the blue-shifted, outflow asymmetry. This 
cloud is apparently experiencing simultaneous
but spatially coordinated, infall and outflow motions of its outer layers. Such 
an alternating spatial pattern may be suggestive of global, 
non-radially symmetric surface oscillations or pulsations of the cloud about
an equilibrium configuration as discussed later in this paper.

\section{Discussion}

\subsection{A Thermally Supported Cloud}

The striking agreement between the observed density distribution in the B68
cloud and that predicted for a Bonnor-Ebert sphere near critical stability,
suggests that thermal pressure is a primary source of support for this cloud
against collapse.  To evaluate this possibility we obtained deep observations
of two optically thin species, \ceio and \ctfs at two locations.  One at the
center (0E 0W) of our mapping grid, near the projected  
center of the cloud, and a second
at a position 36 arc seconds east and 72 arc seconds north of our grid center
(i.e, 36 E 72 N) in the outer regions of the cloud.  We find the linewidths at
the center position (after correcting for instrumental resolution) to be 0.18
$\pm$ 0.01 \kms and 0.15 $\pm$ 0.01 \kms for \ceio and \ctfs, respectively.  At
the outer (36 E 72 N) position these widths are 0.28 $\pm$ 0.01 and 0.22 $\pm$
0.01 \kms, respectively.  These are extremely narrow linewidths and the fact
that the \ctfs lines are narrower than the \ceio lines is suggestive of a 
purely thermal line broadening mechanism.

The linewidth of an optically thin species (x) is given by:

\begin{equation}
\Delta V^2_{\rm x} = 8ln2 k T_k / m_{\rm x} + \Delta V^2_{NT} 
\end{equation}

\ni 
where $m_{\rm x}$ is the mass of the molecular species, x, $\Delta
V_{NT}$ \ is the contribution to the linewidth due to non-thermal motions, 
either random (turbulent) or systematic (i.e., contraction, expansion, rotation)
and $T_k$ is the kinetic temperature of the gas.  In
principle, with observations of two species of differing molecular mass one can
solve the corresponding set of equations (2) and simultaneously determine $T_k$
and $\Delta V_{NT}$.  However, our observations indicate that the CO 
and CS molecules in B68 suffer significant differential depletion 
\citep{bahl02}.  As a
result, these lines likely originate from different volumes of the
cloud. Given that there is also an observed linewidth gradient in this cloud, 
comparison of these linewidths to simultaneously derive the cloud temperature 
and turbulent velocity is probably not a valid procedure.

However, with independent knowledge of the kinetic temperature,
Equation 2 can be used to estimate the relative contribution of 
non-thermal and thermal motions to the observed linewidth
for each species individually, with the additional assumption
of an isothermal cloud. Early 
observations of $^{12}$CO by \citet{mb78} suggested a kinetic
temperature of 11 K, which was consistent with an independent estimate
of $\leq$ 12 K from NH$_3$ (1--1) and (2--2) observations by the same
authors. More recent NH$_3$ observations by \citet{bhrjw95} indicated
a temperature of 16 K, although this estimate may be too high
\citep{tb02}. Analysis of a $^{12}$CO(2--1) spectrum of the central
regions of B68, kindly obtained for us by Dr. Frank Bensch (2002) with the
Gornergrat telescope in Zermatt Switzerland, indicate a kinetic temperature
of T$_k$ = 10.5 $\pm$ 1 K, in excellent agreement with the earlier \citet{mb78}
observations and very recent NH$_3$ observations \citep{hhj02} which came
to our attention after the initial submission of the present paper. 
This is also consistent with the kinetic temperature
derived from observations of \ntwohp \citep{bahl02}.
We therefore adopt this as the kinetic temperature of 
the cloud and from here forward and assume that the cloud is isothermal.

The ratio of thermal to non-thermal (or turbulent) pressure in the cloud is
given by:

\begin{equation}
R_p = {a^2\over\sigma_{NT}^2}
\end{equation}

\ni
where $a$ is the isothermal sound speed and $\sigma_{NT}$ is the 
3-dimensional, rms non-thermal velocity dispersion. For T$_k$ = 10.5
and using Equation 3 we determine  
$\sigma_{NT}$ = 0.104 \kms, for the \ceio data and $\sigma_{NT}$ = 0.086
\kms\ for the \ctfs data. Thus we find 
$R_p = 4-5.$ from these observations, indicating that the
central regions of the B68 cloud map are
clearly dominated by thermal pressure. In the outer region (36E 72N)
of the cloud where the linewidths are larger, and presumably turbulence 
or systematic motions more important, a similar analysis yields $R_p = 1-2$, 
from the observed \ceio and \ctfs linewidths. 
Even in the outer regions, thermal motions are a significant, if not
dominant, source of pressure
for the cloud. These considerations demonstrate that B68 is a thermally
dominated cloud and confirms the inference to this effect
drawn from the analysis of the extinction derived, column density 
profile of the cloud (ALL01).

Finally, additional support for a predominately thermal velocity field is also
provided by examination of the cloud's radial velocity distribution.  As
mentioned earlier, there is a systematic, global variation in the spatial
distribution of line center velocities of the optically thin, single-peaked
\ceio \ and \ntwohp \ lines.  Both these species show well behaved radial
velocity gradients across the cloud.  In Section 3.2 we
used a $\chi^2$ fitting procedure to
fit a 2-dimensional plane to each velocity field and derive the magnitudes
and directions of their velocity gradients.
If we remove the derived 2-dimensional radial velocity gradient from
the data, the residual line center velocities at neighboring positions 
can be compared and used to measure the magnitude of any relative random
bulk motion of the molecular gas along a given line-of-sight across the cloud. 
For example, for a purely thermal cloud we would expect the difference in residual 
velocities
to be small, while for a turbulent cloud these differences 
would to some extent reflect the magnitude of the turbulent velocity dispersion. 

After
removing the derived 2-dimensional velocity gradient from the observed
\ceio velocity field, we computed the velocity difference ($\delta V_{nn}$)
between the nearest neighbor locations in the map. We found the average
velocity difference between nearest neighbors to be 
$$<|\delta V_{nn}|> = 11.16 \pm 0.18 \ \ {\rm m \ s^{-1}}$$

\ni
for observations separated by 12 arc sec and 15.01 $\pm$ 0.19 m s$^{-1}$
for points separated by 24 arc sec. These velocity differences are
considerably smaller
than either the typical one-dimensional rms turbulent velocity
dispersion ($\sigma$ = 60 - 100 m s$^{-1}$), or the 
\ceio linewidths  ($\sim$ 180 - 208 m s$^{-1}$), or the expected sound
speed in the cloud ($\sim$ 200 m s$^{-1}$), which themselves are of comparable 
magnitude. We obtain similar values of 7.34 $\pm$ 0.61 m s$^{-1}$ and
15.2 $\pm$ 0.63 m s$^{-1}$ \ for \ntwohp at 12 and 24 arc sec spacings, 
respectively.  The small values of these nearest neighbor residual velocities
are clearly consistent with and suggest a thermally dominated velocity 
field for this cloud.

\subsection{Radial Velocity Distribution}

Examination of the distribution of the line center velocities of 
two molecules, \ceio and \ntwohpp, which are single-peaked 
reveal clear, well behaved and systematic velocity 
gradients across the cloud. In both maps the iso-velocity
contours are largely parallel to each other and so
suggest solid body rotation of the cloud around a north-south
axis. The ratio of rotational kinetic energy to gravitational energy,
$\beta$, is often used to quantify the dynamical importance of
rotation for cloud stability and is given by \citep{gbfm93}:

$$\beta = {1\over 2} {p\over q} {\omega^2 R^3\over GM}$$

\ni
Here, $M$ and $R$ are the cloud mass and radius, respectively,
and p and q are defined so that the moment of intertia and
the gravitational potential energy are given by $pMR^2$, and
$qGM^2/R$, respectively. For a stable spheroid the virial theorem
suggests that $0 \leq \beta \leq 0.5$.  For the parameters of B68,
$M$= 1.6 \msun, $R$ =  1.5 $\times$ 10$^{17}$ cm, and $\omega =
dv/ds/sin(i)$, where $i$ is the inclination of the cloud to the
line-of-sight, we find $\beta = 0.18\ p/q\ sin^{-2}(i)$ for the 
measured \ceio rotation rate. The
ratio p/q is likely to have a value somewhat less than 1, for example,  
for a sphere with an r$^{-2}$ density profile, $p/q$ = 0.22, while
for a constant density sphere, $p/q$ = 0.67.  
Because the lines of constant radial velocity 
are so nearly parallel across the face of the cloud, $i$ must be much closer
to  90$^{\rm o}$ than to 0$^{\rm o}$. Assuming $sin(i) = 1$ and 
$p/q$ = 0.22 we estimate $\beta_{\rm B68}$ = 0.04. Thus, the rotational
kinetic energy is only a few percent of the gravitational potential energy
and the contribution of 
rotation to the overall dynamical stability of the B68 cloud is 
not very significant. This value of $\beta$ is typical of that estimated
for numerous other molecular cloud cores by \citet{gbfm93}. This value
is also consistent with the Bonnor-Ebert fit to the cloud density distribution
which suggests that the cloud is primarily supported by thermal pressure 
\citep{all01}.

Because \ceio and \ntwohp emission lines are formed in different volumes of the
cloud due to differential depletion \citep{bahl02}, their velocity fields also must
characterize the gas in different cloud volumes.  In particular, \ntwohp probes
more interior regions than \ceio.  In this respect, it is interesting that the
rotational velocity gradients derived from the two tracers appear to differ.  
If the cloud
were rotating like a solid body, both tracers should exhibit the same projected
radial velocity gradient. Both the magnitude of the rotational velocity gradient
and its projected direction on the sky differ more than the quoted 
errors in these quantities.
The magnitude of the rotational gradient derived 
from the \ceio data is a factor of 1.3 greater than that derived from the 
\ntwohp data. This appears to indicate that the cloud is
experiencing differential rotation, with the inner regions rotating somewhat
more slowly than the outer regions. Although higher velocity resolution 
mapping obsevations of \ntwohp would be required to confirm this interpretation
of the data. Nontheless, the radial profile of angular velocity for 
this cloud differs from the inward spin up one would expect
for the more Keplerian-like motion of an isothermal sphere in near hydrostatic
equilibrium \citep{knmh87,t78} and the angular velocity profile
predicted for an evolving magnetized protostellar core (Basu \& Mouschovias
1995a, b). However, in the absence of significant ambipolar diffusion
in the cloud core, one might expect magnetic breaking 
to force co-rotation between the inner and outer parts of the cloud.
This would require the existence of a 
magnetic field which permeates the cloud and is coupled to the gas.
The lack of a significant non-thermal component to the observed line 
widths in the central regions would seem to preclude a significant turbulent 
magnetic field. Although a rigid, static field could be
present, it might be relatively weak, given the close balance between
gravitation and thermal pressure required by the  agreement of 
the cloud's density profile and the predictions of a near critical
thermally supported Bonnor-Ebert sphere \citep{all01}. Thus, it is presently
unclear whether or not magnetic breaking or some other cause is
responsible for the lack of increased rotation in the central regions 
of the cloud.

\subsection{Self-Reversed Profiles: Non-Radially Symmetric Pulsation?}

The CS self-reversal traces the motions of gas in the outer layers of the cloud.
The systematic spatial segregation of the line profile asymmetry across 
the B68 cloud is intriguing and is not what is expected from a simple
contraction or expansion of the outer cloud layers. Instead the observations appear
to indicate that gas in the central regions of the map is contracting while 
{\it simultaneously} the gas in the outer regions of the cloud map
is expanding. This description is somewhat
of an oversimplification of the true situation since the red-shifted asymmetry
extends to the gas in the southeast tail of the cloud, where the CS emission 
is the strongest. Nonetheless the overall pattern across the main body
of the cloud is suggestive of some type of organized or co-ordinated global mass
motion. The observed inflow and outflow velocities are subsonic and therefore
suggest a small perturbation on the overall dynamical state of the cloud. 
Indeed, the observed pattern over the main portion of the cloud is similar 
to that which would be expected from low frequency spatial oscillations of 
the outer cloud layers around some equilibrium 
dynamical state.  Small amplitude oscillations of a spherical surface can usually
be described in terms of spherical harmonic functions (e.g., Tassoul 1978). 
The real part of the
radial component of these functions can produce organized patterns of
alternating infalling and outflowing motions across the surface of an oscillating
or pulsating sphere, as for example, has been observed on the surface of the sun.
In this case the velocity field may be described as:
$$V(\theta,\phi;t) = \sqrt{4\pi} (-1)^m c_{lm} P^m_l(\cos \theta) 
cos(m\phi - \omega_0t)$$
\noindent
where $P^m_l(cos \theta)$ is the Legendre function, $c_{lm}$, a normalization
constant, and $\omega_0$ is the oscillation frequency (e.g., Christensen-Dalsgaard 1998).
The oscillation frequency is likely to be proportional to the cloud density, 
that is, $\omega_0^2 \propto
 \rho$, with the constant of proportionality depending on the physical nature
of the restoring force for the oscillation (e.g., Tassoul 1978).
The number of surface sectors which are either infalling or outflowing is 
related to the modal values, $l$ and $m$, of the corresponding Legendre function. 
This is illustrated in Figure 8 which was kindly
provided to us by Clem Pryke of the University of Chicago. This figure
displays the real parts of the spherical harmonic functions, 
$Y^m_l(\theta, \phi) = (-1)^m c_{lm}P^m_l(cos \theta) exp(im\phi)$,
for a series of low order modes. The red and blue colors in Figure 8 can be 
considered to respectively map the inward and outward moving sectors of an 
oscillating spherical surface. To an external observer, the corresponding velocity 
field projected along the line-of-sight (for a fixed inclination of the sphere) 
would maintain a similar spatial distribution or pattern of red and blue motions, 
although the magnitude of the observed velocites would decrease towards the edges of 
the observed disk where tangential motions would be the greatest. 
The pattern observed on B68, with the central region
moving in one direction (inward) and the surrounding regions moving in
another (outward) is very similar to that expected for an $l$ = 2, $m$ = 2 mode
of oscillation, with the polar axis of the sphere corresponding to the 
rotational axis of the cloud. In this case the spherical cloud would be 
divided into four equal sectors with the sectors on the opposite side of 
the sphere moving in the same direction. This pattern corresponds to a quadrupole
oscillation of the outer layers of the cloud.

From our data it is difficult to determine the extent to which such oscillatory
motions penetrate below the cloud's surface.  The relatively long wavelength of the
$l$ = 2 perturbation suggests that deeper cloud layers may also be in oscillation.
It is even possible that radial modes of oscillation are present in the interior of
the cloud along with the non-radial surface modes.  To assess the depth to which
surface oscillations may extend or whether radial modes of oscillation are present
requires the ability to measure inflow/outflow motions of gas in regions deeper in
the cloud than traced by CS emission.  This is a very difficult task since it
requires the existence of a tracer which both preferentially traces inner cloud
material and simultaneously is optically thick.  In this context it is intriguing
that the high spectral resolution profile of N$_2$H$^+$ obtained at the center of
the cloud map (Figure 2) appears to be asymmetric.  As mentioned earlier, the
observed N$_2$H$^+$ emission has modest optical depth and thus the asymmetry in the
profile shape could be the result of self-absorption.  Therefore, because \ntwohp
is considerably less depleted in the core of the cloud, the asymmetry in its
profile may be a useful probe of gas motions interior to those traced by the CS
emission.  The fact that the N$_2$H$^+$ asymmetry has the opposite sense of the
asymmetry present in the corresponding CS line (along the same line-of-sight)
suggests that the \ntwohp emitting gas in the inner regions is moving in a
direction opposite that of the CS emitting gas in the outer regions.  This, in
turn, may suggest the presence of a radial mode of oscillation with a node located
somewhere between the CS envelope and N$_2$H$^+$ core of the cloud.  However, this
inference of the presence of radial modes of oscillation is not by any means
definitive, since it depends on the interpretation of a single high-resolution
spectrum of \ntwohp obtained at the center of the cloud map.  Mapping of the
\ntwohp asymmetry across the cloud along with more detailed radiative transfer
calculations which model individual hyperfine components to better constrain the
line opacity would be very useful for confirming the existence of such internal
cloud motions.

The Barnard 68 cloud appears to be embedded in the Loop I supernova super bubble
\citep{all01,all02}.  Its derived surface pressure is an order of magnitude
greater than that of the general ISM but comparable to that derived for the Loop
I super bubble from x ray observations (Breitschwerdt, Freyberg \& Egger, 2000).
Therefore, the cloud likely interacted with the shock of a supernova remnant
sometime during the not too distant past.  We speculate that this external
interaction with the supernova provided the peturbation that set at least the
outer layers of the cloud into oscillation.  Although the interaction did not
de-stablize the cloud, it may have excited other, higher mode oscillations which
have since damped out.  One possibility is that B 68 was a dense core region of a
much more massive cloud complex.  In this case it would have been surrounded by a
lower density, cold molecular envelope, the weight of which provided a similar
confining external pressure to that characterizing the cloud now.  This envelope
would have been stripped with the passage of the supernova shock, leaving behind
the denser core region which then came into a new pressure equilibrium with the
hot gas in the supernova shell.  The oscillations we now observe might then be
the remaining signature of that physical interaction.

\section{Summary}

We have used the 30-m IRAM millimeter telescope to obtain sensitive,
high angular resolution observations of \ceio, C$^{32}$S, \ctfs and
\ntwohp in the dark cloud Barnard 68. Our primary results can be summarized as 
follows:

\noindent
1) The \ceio, \ctfs and \ntwohp lines are characterized by very narrow
linewidths in the central regions of the cloud map consistent with nearly
pure thermal line broadening. We derive the ratio of thermal to non-thermal
pressure to be R$_p$ = 4 -- 5 from analysis of the \ceio and \ctfs 
line profiles at the cloud center. We further demonstrate the rotational
energy of the cloud to be only a few \% of the gravitational
potential energy.  B68 appears to be a thermally dominated
cloud, where thermal pressure is the dominant 
source of support against gravity. This confirms inferences to this effect
resulting from the close correspondence of the measured density 
distribution of the cloud with predictions for a critically stable
Bonnor-Ebert sphere \citep{all01,all02}.

\noindent
2) The linewidths of individual molecular species are found to 
systematically increase with distance from the projected center
of the cloud. In
the outermost regions of the cloud we find R$_p$ to decrease 
to between 1 and 2, indicating an increasing role of non-thermal
or turbulent motions to the overall cloud pressure in those 
regions.

\noindent
3) The spatial distributions of line center radial velocities of \ceio and
\ntwohp display systematic, linear velocity gradients individually
indicative of solid body rotation. However both the direction and magnitude
of the gradients observed in the two species differ, suggesting the presence
of differential rotation in the cloud. The region traced by \ntwohp
emission appears to be rotating 1.3 times slower than the region 
traced by \ceio emission. Since the latter species is known to be
significantly depleted in the interior regions of the cloud this may
suggest that the inner regions are rotating more slowly than 
the outer regions, contrary to expectations. However, higher velocity
resolution spectral observations of \ntwohp are required to confirm the presence
of such differential rotation. 

\noindent 
4) The C$^{32}$S lines are found to be optically thick, saturated and
self-reversed over nearly the entire surface of the B68 cloud.  Moreover, the
self-reversed profiles are asymmetric with both red-shifted asymmetric (infall) and
blue-shifted asymmetric (outflow) profiles found in the cloud.  The distribution
of line profile asymmetry displays a systematic spatially alternating pattern of inflow
and outflow motion of the cold outer cloud material.  The central regions of the cloud map
are characterized by infall motions, while the surrounding regions are
primarily characterized by outflow motions.  This global pattern of alternating infall
and outflow motion aligned roughly along the cloud rotational axis is suggestive of a 
small amplitude, low order mode, non-radial pulsation or oscillation of 
the outer surface layers of the cloud around an
equilibrium configuration. In addition, the asymmetry revealed in a high resolution 
spectrum of \ntwohp emission obtained at the center of the cloud map may further 
suggest the presence of a radial mode of oscillation with a node located between
the outer CS emitting and inner \ntwohp emitting regions of the cloud. 
Such oscillations might be the signature of an earlier
interaction of the cloud with a passing supernova shock.

\acknowledgments

We are grateful to Frank Bensch for acquiring a spectrum of $^{12}$CO
at the center of B68. We thank Shantanu Basu, Sylvain Kozennik, Robert Noyes, 
Ergenya Shkolnik and Frank Shu for useful discussions. We thank an anonymous
referee for suggestions which helped clarify results presented in the paper.
This research was supported by NASA Origins Grant NAG 5-9520.



%\appendix

%\section{Appendicial material}

\clearpage


\begin{figure}
\plotone{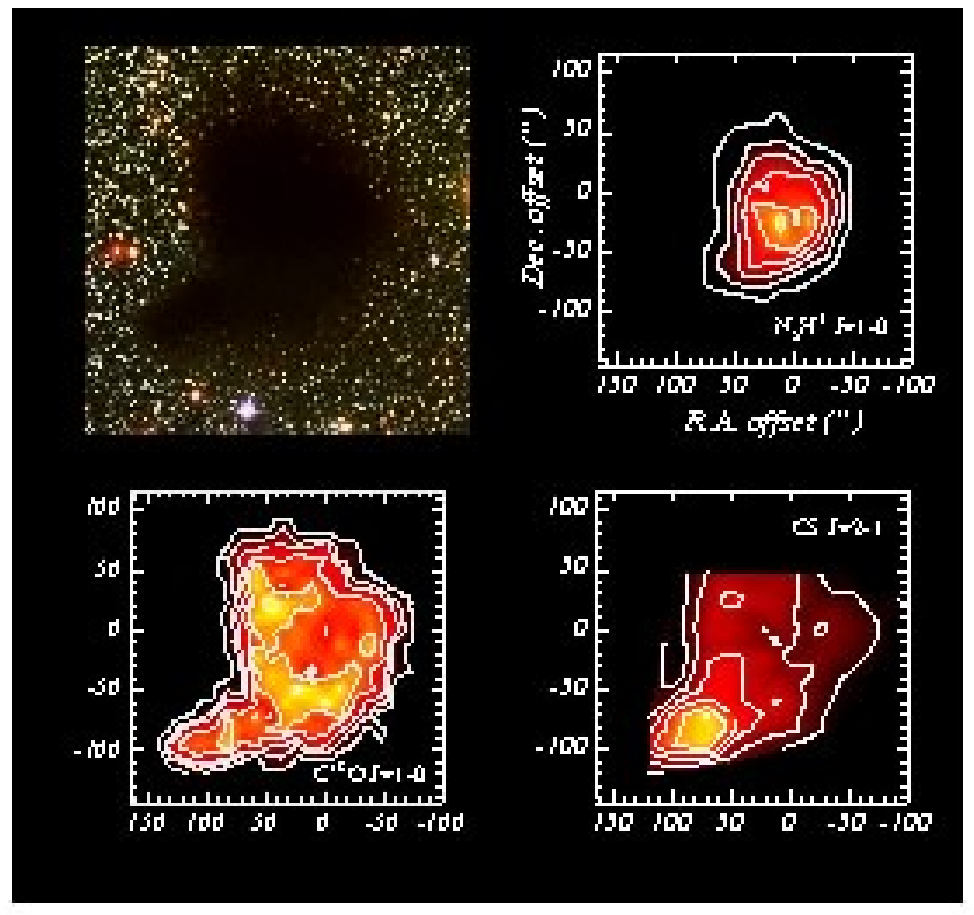}
\caption{A deep optical image of the dark globule
Barnard 68 (Alves, Lada \& Lada 2001) along
with contour maps of integrated intensity from molecular emission 
lines of N$_2$H$^+$, C$^{18}$O, and CS. Differences in appearance 
of the molecular line maps are primarily due to the effect of 
increasing molecular depletion onto grains with \ntwohp being the 
least depleted and CS the most depleted of these species in the 
cloud core (Bergin et al. 2002). 
Contour levels begin at 0.3, 0.2, 0.15 K km s$^{-1}$ and 
increase in intervals of 0.2, 0.1, and 0.10 K km s$^{-1}$ 
for N$_2$H$^+$, C$^{18}$O, and CS, respectively. The poor quality
of this image is a result of the size limitations imposed by 
astro-ph; the full resolution figure can be accessed elsewhere:
%\url{http://cfa-www.harvard.edu/~ebergin/pubs_html/b68_vel.html}
\label{fig1}}
\noindent
\url{http://cfa-www.harvard.edu/~ebergin/pubs\_html/b68\_vel.html.}
\end{figure}

\clearpage 

\begin{figure}
\plotone{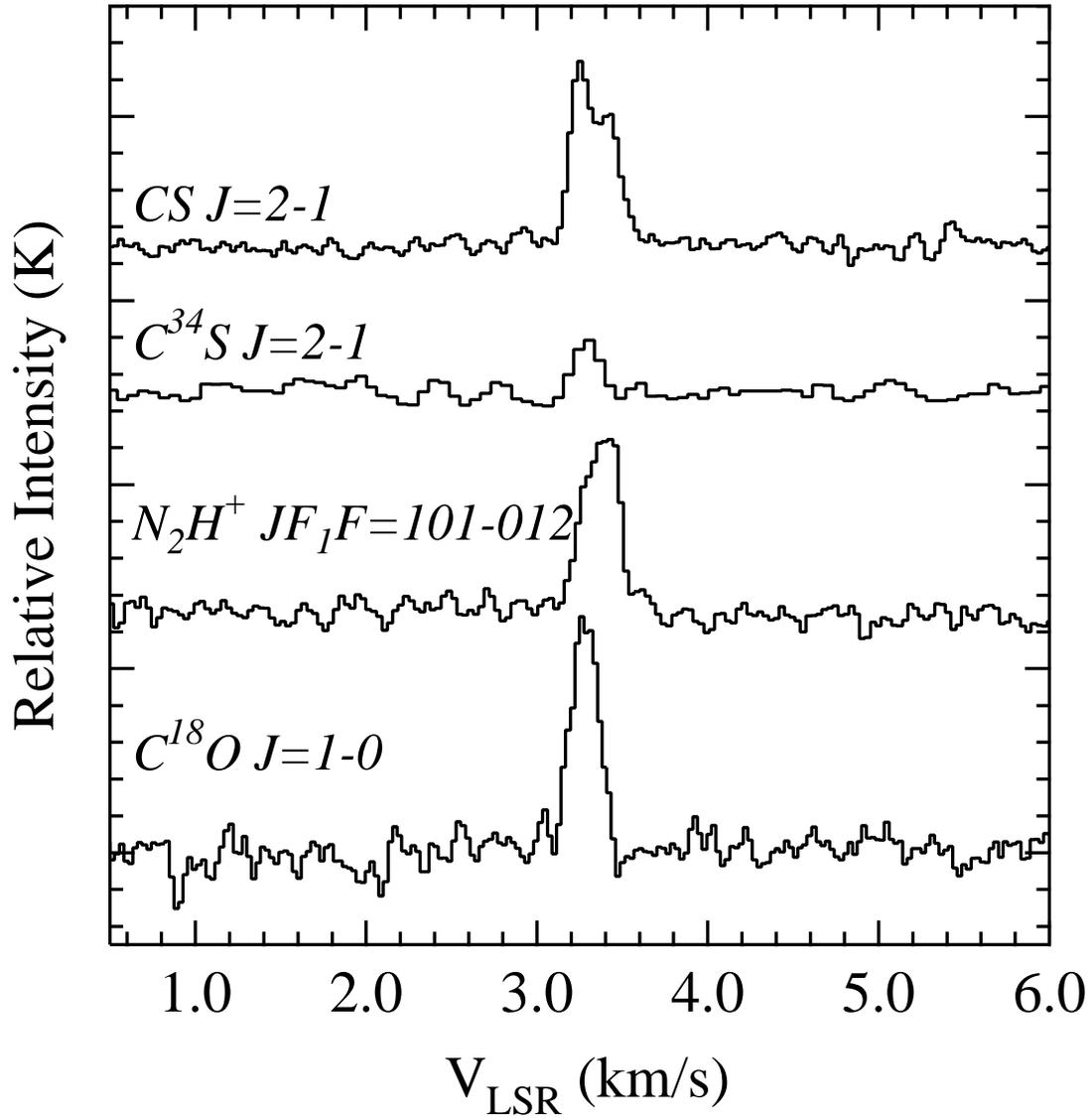}
\caption{ Spectra of CS, \ctfs, \ntwohp and \ceio obtained at the 
reference position (0,0) near the center of the globule. The 
\ntwohp spectrum at the reference position was obtained at a significantly
higher velocity resolution (0.03 \kms) than that (0.137 \kms) which 
characterized all other positions in the cloud. \label{fig2}}
\end{figure}

\clearpage

\begin{figure}
\plotone{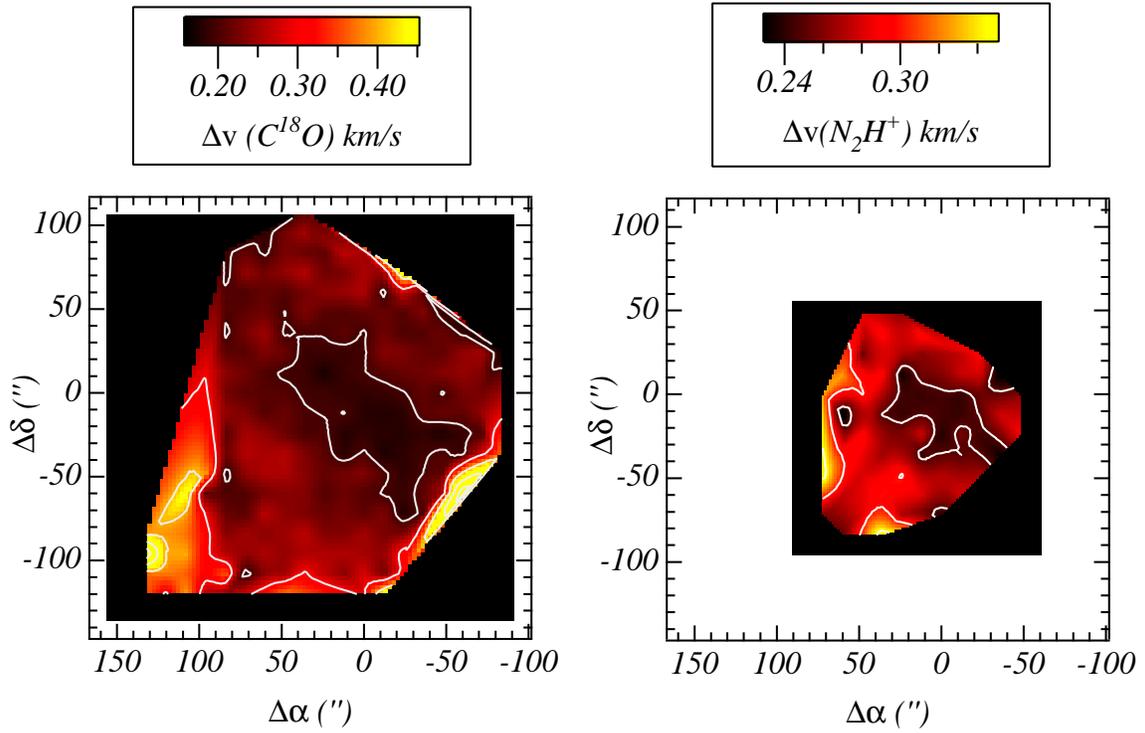}
\caption{Contour maps of the distribution of linewidth,
for C$^{18}$O and N$_2$H$^+$ emission in the B68 cloud. 
The linewidths are generally larger in the outer
regions of the cloud in both tracers. \label{fig3}}
\end{figure}

\clearpage

\begin{figure}
\plotone{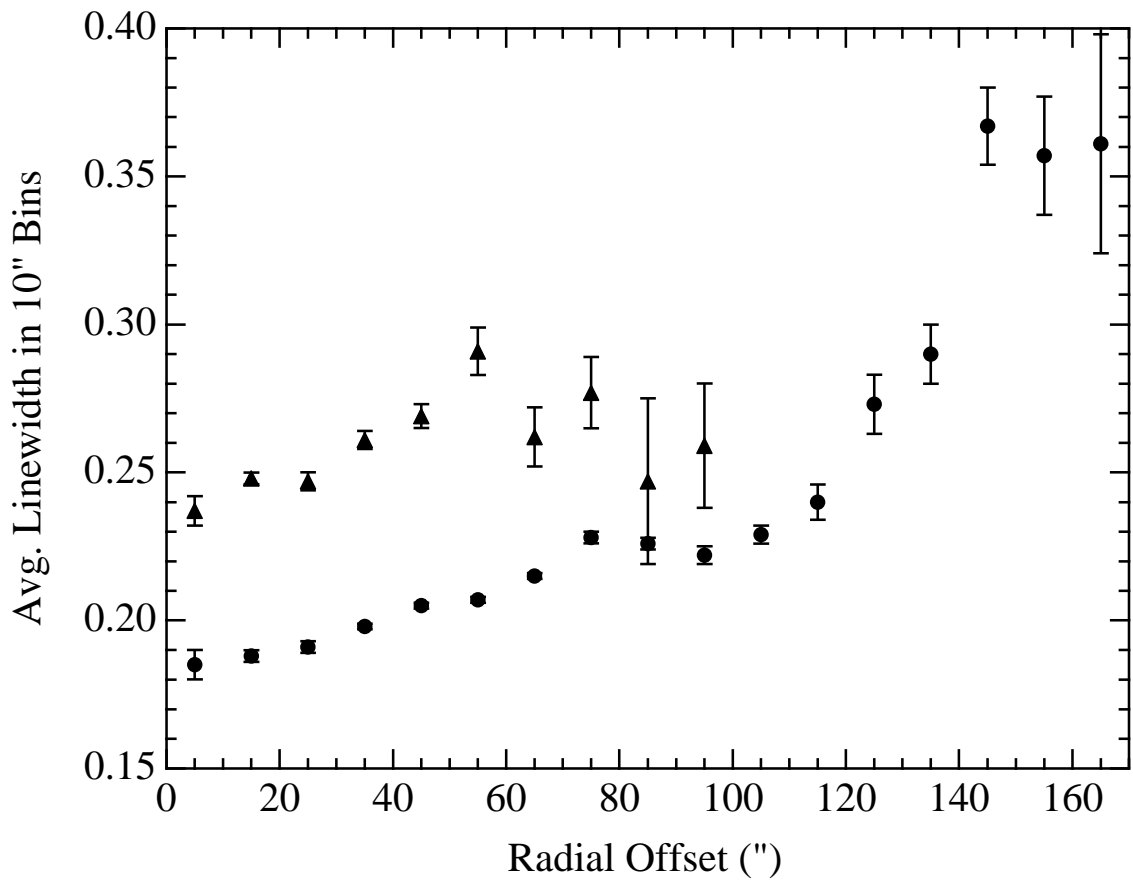}
\caption{The radially averaged variation of N$_2$H$^+$ (triangles) and 
C$^{18}$O (circles) linewidths in the B68 cloud. A trend for
the linewidth to increase with increasing distance from the cloud
center is evident in the inner regions of the cloud for both molecular
tracers. The linewidths of both species
appear to be constant or decline slightly between 60 - 100 arc sec
from the cloud center. At greater distances, where N$_2$H$^+$ emission is 
absent, the C$^{18}$O linewidths appear to increase again. \label{fig4}}
\end{figure}

\clearpage

\begin{figure}
\plotone{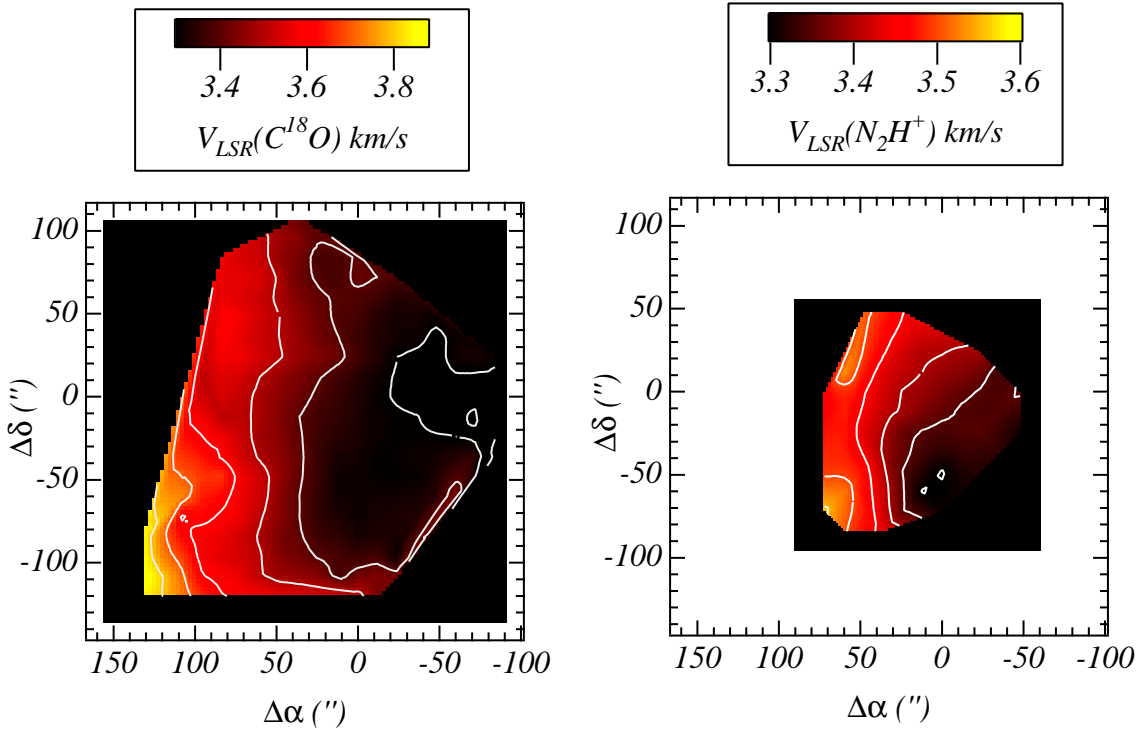}
\caption{Contour maps of the distribution of the peak or line center radial velocity,
V$_{\rm LSR}$, for C$^{18}$O and N$_2$H$^+$ in the B68 cloud. The maps show
a clear and systematic east-west velocity gradient across the cloud in both
tracers. However the magnitude of the N$_2$H$^+$ velocity gradient is somewhat
smaller than that of CO. This difference may suggest differential rotation of the cloud since
the CO molecule is significantly more depleted in the central cloud regions
than N$_2$H$^+$ (Bergin et al. 2001).
\label{fig5}}
\end{figure}

\clearpage 

\begin{figure}
%\plotone{specplot_potrait.eps}
\plotone{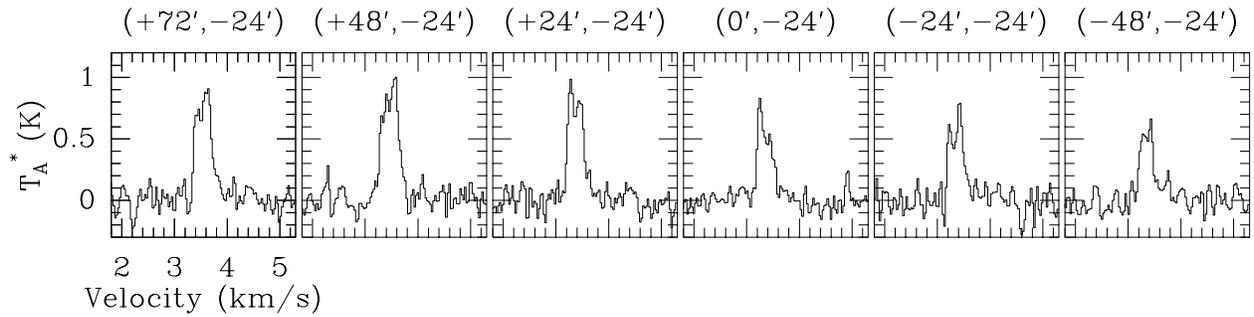}
\caption{A series of CS (J = 2--1) spectra along a cut through the
cloud obtained at a constant declination of 24 arc seconds south of the 
nominal reference position. The profiles are all optically thick and 
exhibit clear asymmetric, self-reversed (multiple-peaked) structure. 
The sense of 
the self-reversed asymmetry alternates in a systematic spatial pattern 
from blue-shifted to red-shifted and back to 
blue-shifted as one proceeds from east to west across the cloud. \label{fig6}}
\end{figure}

\clearpage 

\begin{figure}
%\plotone{b68_csabsvel_rev.EPS}
\plotone{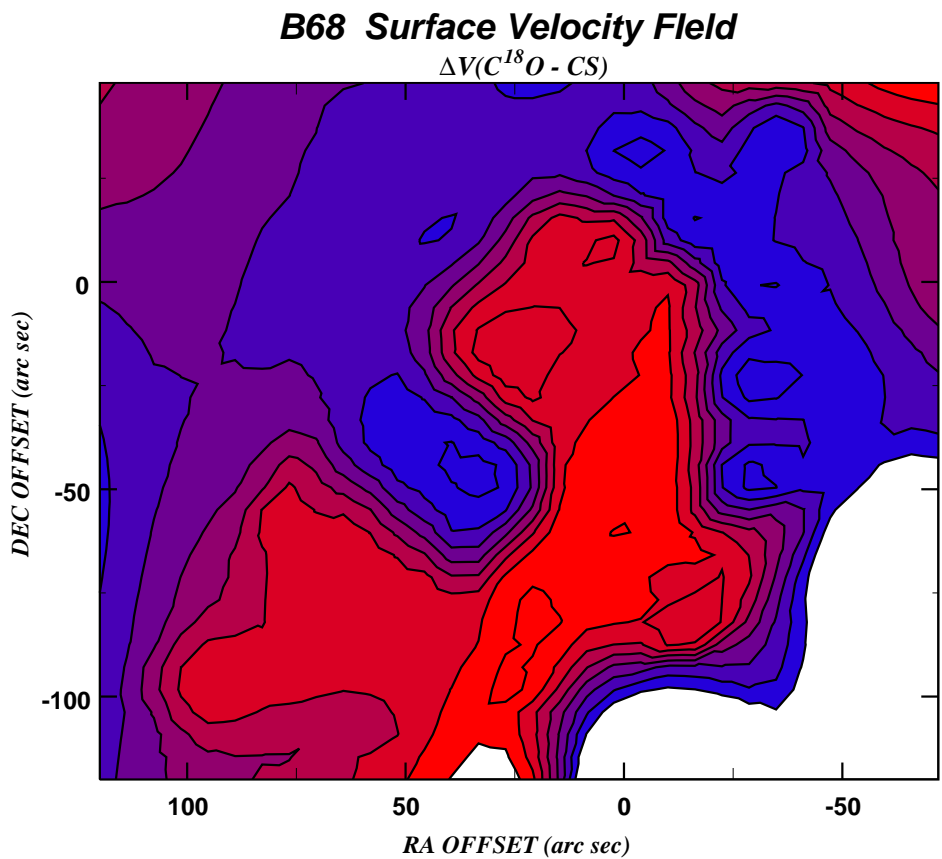}
\caption{Contour map of the distribution of $\delta V$, the velocity difference
between the peak of (optically thin) C$^{18}$O emission and
the strongest peak of (optically thick) CS emission at each position in the 
cloud. A positive difference is coded red and indicates red-shifted or infall motions 
of the CS gas (relative to CO and the systemic velocity of the cloud)
while a negative difference (coded blue) indicates
blue-shifted and outflow motions of the CS gas. There is a systematic, 
spatially alternating, pattern of blue and red-shifted gas motion across 
the cloud. The contours begin at $-$0.16 km s$^{-1}$ and increase in
intervals of 0.04 km s$^{-1}$ to a maximum of +0.12 km s$^{-1}$ \label{fig7}}
\end{figure}

\clearpage 

\begin{figure}
%\plotone{static_poles.ps}
\plotone{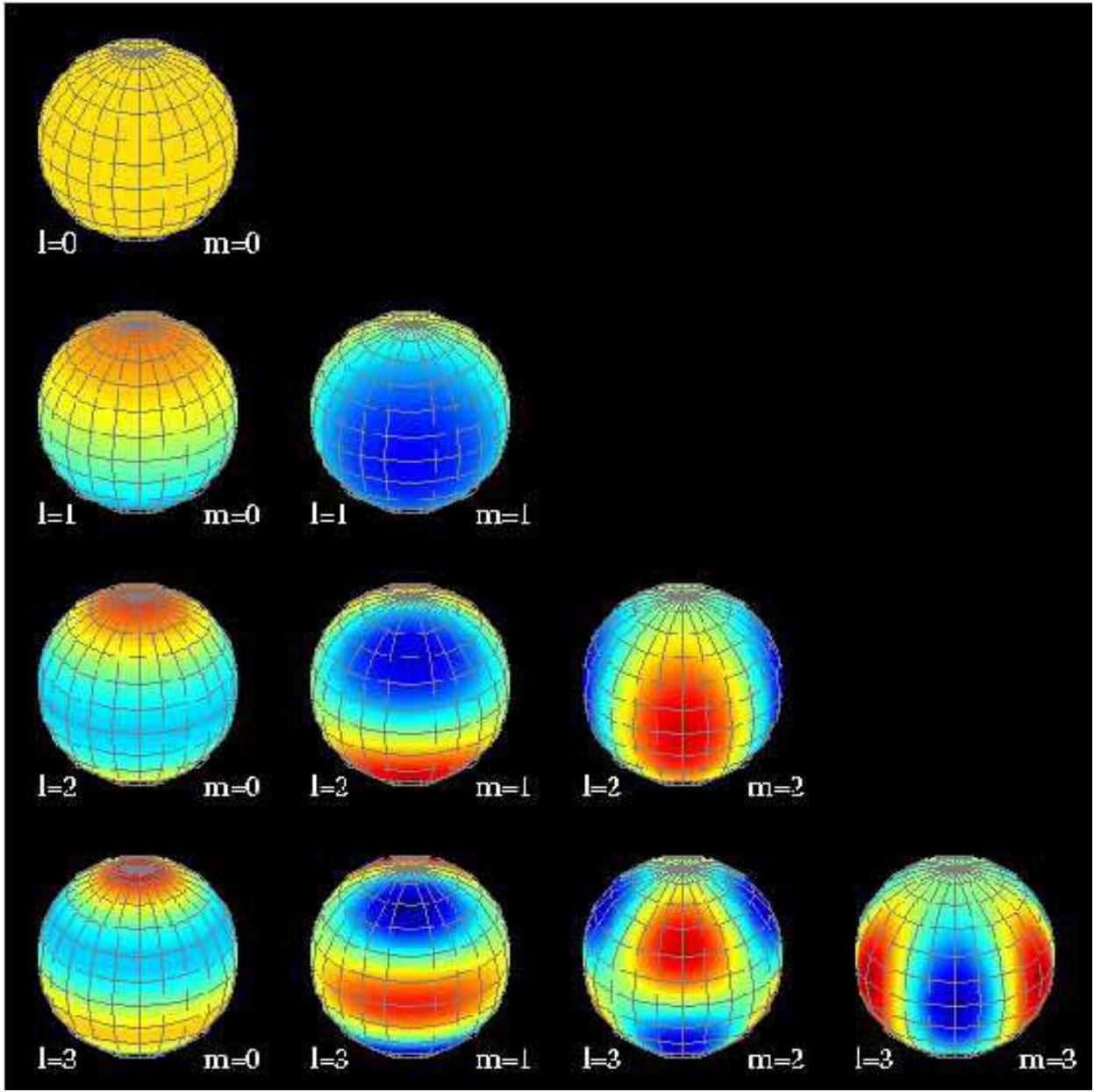}
\caption{False color intensity maps of the real part of a series
of low order spherical harmonics. Visual inspection
suggests that the l = 2, m = 2 mode appears to 
correspond to the distribution of $\delta V$ for
B68 which is displayed in Figure 7. (Figure courtesy of
Dr. Clem Pryke, University of Chicago).
\label{fig8}}
\end{figure}

\clearpage

%% If you are not including electonic art with your submission, you may
%% mark up your captions using the \figcaption command. See the 
%% No more than seven \figcaption commands are allowed per page, 
%% so if you have more than seven captions, insert a \clearpage 

%% The following command ends your manuscript. LaTeX will ignore any text


\begin{thebibliography}{}

 
\bibitem[Alves, Lada \& Lada(2001)]{all01} Alves, J., Lada, C. J., \&
   Lada, E.A. 2001 Nature, 409, 159.

\bibitem[Alves, Lada \& Lada(2002)]{all02} Alves, J., Lada, C. J. \&
  Lada, E. A. 2002, in preparation.

\bibitem[Basu \& Mouschovias (1995a)]{bm95a} Basu, S. \& Mouschovias, T. C.
1995a, ApJ., 452, 386.
 
\bibitem[Basu \& Mouschovias (1995b)]{bm95b} Basu, S. \& Mouschovias, T. C.
1995b, ApJ., 453, 271. 
 
\bibitem[Bergin et al.(2002)]{bahl02} Bergin, E. A., Alves, J. A., Huard,
T. L. \& Lada, C. J. 2002, ApJL, 570, L101.

\bibitem[Bourke(2002)]{tb02} Bourke, T., 2002, personal communication.

\bibitem[Bourke et al.(1995)]{bhrjw95} Bourke, T., Hyland, H., Robinson, G,
James, S. \& Wright, C. 1995, MNRAS, 276, 1067.

\bibitem[Breitschwerdt et al. (2000)]{bfe00} Breitschwerdt, D, Freyberg, M.,
\& Egger, R. 2000, A\&A, 3114, 258.

\bibitem[Christensen-Dalsgaard (1998)]{cd98} Christensen-Dalsgaard, J. 1998,
Lecture Notes on Stellar Oscillations, 4th Edition, 
URL: http//www.obs.aau.dk/$\sim$jcd/oscilnotes/

\bibitem[Hotzel Harju \& Juvela (2002)]{hhj02}
Hotzel, S., Harju, J. \& Juvela, M. 2002, A\&A, in press.

\bibitem[Goodman et al.(1993)]{gbfm93} Goodman, A. A., Benson, P. J., 
Fuller, G. A., \& Myers, P. C. 1993, ApJ, 406, 528.

\bibitem[Juvela(1998)]{juvela98} Juvela, M. 1998, A\&A, 338, 723.

\bibitem[Kiguchi et al.(1987)]{knmh87} Kiguchi, M., Narita, S., Miyama, S. M., 
Hayashi, C. 1987, ApJ, 317, 830.
 
\bibitem[Martin \& Barrett(1978)]{mb78} Martin, R. N., \& Barrett,
A. H. 1978, ApJS, 221, 124.

\bibitem[Myers et al.(1996)]{mmtww96} Myers, P.C., Mardones, D., Tafalla, M.,
Williams, J.P., \& Wilner, D.J. 1996, ApJL, 465, 133.

\bibitem[Padoan, Jones \& Nordlund(1997)]{pbn97} Padoan, P., Jones, B.T.,
     \& Nordlund, A.P. 1997, ApJ, 474, 730.
     
\bibitem[Tassoul(1978)]{t78} Tassoul, J.-L. 1978, Theory of Rotating Stars 
(Princeton University Press, Princeton). 

\bibitem[Walker et al.(1986)]{wlymw86} Walker, C.K., Lada, C.J., Young, E.T., 
Maloney, P.R., \& Wilking, B.A. 1986, ApJL, 309, 47.

\end{thebibliography}
\end{document}